\documentclass{elsart3-1}

\usepackage{amssymb}
\usepackage{amsmath}
\usepackage[numbers]{natbib}
\usepackage[english,francais]{babel}
\newcommand{\R}{\mathbb{R}}
\newcommand{\E}{\mathbb{E}}

\newcommand{\Tr}{\mathrm{Tr}}
\newcommand{\Xu}{X_{\bf u}}
\newcommand{\Cu}{C_{\bf u}}
\newcommand{\Xtu}{X_{\sim\bf u}}
\newcommand{\Ctu}{C_{\sim\bf u}}
\newcommand{\Cutu}{C_{\bf u,\sim\bf u}}

\newtheorem{theorem}{Theorem}[section]
\newtheorem{lemma}[theorem]{Lemma}
\newtheorem{e-proposition}[theorem]{Proposition}

\newtheorem{e-definition}[theorem]{Definition\rm}

\renewcommand{\theequation}{\arabic{equation}}
\def\og{\leavevmode\raise.3ex\hbox{$\scriptscriptstyle\langle\!\langle$~}}
\def\fg{\leavevmode\raise.3ex\hbox{~$\!\scriptscriptstyle\,\rangle\!\rangle$}}

\journal{the Acad\'emie des sciences}
\begin{document}
% place in the next line the header (rubrique) chosen for your article,
% if you know it (you can also have 2, format : Header1/Header2
\centerline{}
\begin{frontmatter}


% \author{Name\thanksref{label2}}
\selectlanguage{english}
\title{Sensitivity indices for multivariate outputs}

% The [label1] can be suppressed if there is only one address for all authors

\selectlanguage{english}
\author[authorlabel1]{Fabrice Gamboa},
\ead{fabrice.gamboa@math.univ-toulouse.fr}
\author[authorlabel2]{Alexandre Janon}
\ead{alexandre.janon@imag.fr}
\author[authorlabel1]{Thierry Klein}
\ead{thierry.klein@math.univ-toulouse.fr}
\author[authorlabel1]{Agn\`es Lagnoux}
\ead{lagnoux@univ-tlse2.fr}
%\author[authorlabel5]{Cl\'ementine Prieur}
%\ead{clementine.prieur@imag.fr}
\address[authorlabel1]{Institut Math\'ematique de Toulouse, 118 route de Narbonne, 31062 Toulouse Cedex.}
\address[authorlabel2]{Laboratoire SAF, ISFA Universit\'e Lyon 1, 50 avenue Tony Garnier, 69007 Lyon}

% If you know the dates of reception, and acceptation you can put them now;
%  idem the name of the person presenting the Note

\medskip
\begin{center}
{\small Received *****; accepted after revision +++++\\
Presented by £££££}
\end{center}

\begin{abstract}
\selectlanguage{english}
% Text of abstract in English
We define and study a generalization of Sobol sensitivity indices for the case of a vector output. 
{\it To cite this article: F. Gamboa, A. Janon, T. Klein, A. Lagnoux, C. R. Acad. Sci. Paris, Ser. xx xxx (2013).}

\vskip 0.5\baselineskip

\selectlanguage{francais}
% Text of abstract in French
\noindent{\bf R\'esum\'e} \vskip 0.5\baselineskip \noindent
{\bf Indices de sensibilit\'e pour sorties multivari\'ees. }
Nous d\'efinissons et \'etudions une g\'en\'eralisation des indices de Sobol pour des sorties vectorielles.
{\it Pour citer cet article~: F. Gamboa, A. Janon, T. Klein, A. Lagnoux, C. R. Acad. Sci. Paris, Ser. xx xxx (2013). }

\end{abstract}
\end{frontmatter}

% now the Version française abrégée, if it exists
%\selectlanguage{francais}
%\section*{Version fran\c{c}aise abr\'eg\'ee}
% Text of your Version française abrégée here.
% Note you do not need to repeat here equations that you use in the
% main text - for example 'voir (3)' is quite acceptable.

\selectlanguage{english}
% main text
\section{Introduction}
\label{}
Many mathematical models encountered in applied sciences involve a large number of poorly-known parameters as inputs. It is important for the practitioner to assess the impact of this uncertainty on the model output. An aspect of this assessment is sensitivity analysis, which aims to identify the most sensitive parameters. In other words, parameters that have the largest influence on the output. In global stochastic sensitivity analysis, the input variables are assumed to be  independent random variables.  Their probability distributions  account for the practitioner's belief about the input uncertainty. This turns the model output into a random variable.

When the output is scalar, using the so-called Hoeffding decomposition \cite{van2000asymptotic}, its total variance can be split down into different partial variances. Each of these partial variances measures the uncertainty on the output induced by the corresponding input variable. By considering the ratio of each partial variance to the total variance, we obtain a measure of importance for each input variable called the \emph{Sobol index} or \emph{sensitivity index} of the variable \cite{sobol1993}; the most sensitive parameters can then be identified and ranked as the parameters with the largest Sobol indices. 

Generalization of the Sobol index for multivariate (vector) outputs has been considered in \cite{lamboni2011multivariate} in an empirical way. In this note, we consider and study a new generalization of Sobol indices for vector outputs. These indices stem from an Hoeffding decomposition and satisfy natural invariance properties. In this note, we define the new sensitivity indices, examine some of their properties and show why they are natural. We also study a Monte-Carlo estimator of these indices, as in practice the exact values are not explicitly computable.

\section{Definitions and Properties}
\subsection{Preliminaries}
We denote by $X_1, \ldots, X_p$ the input random variables defined on some probability space $(\Omega,\mathbb P)$, and by $Y$ the output: $ Y = f(X_1, \ldots, X_p) $, where $f: \R^p \rightarrow \R^k$ ($p,k$ are integers). We suppose that $X_1,\ldots,X_p$ are independent, that $Y \in L^2(\Omega, \R^k)$, and that the covariance matrix of $Y$ is positive definite.

For any non-empty $r$-subset $\bf u$ of $\{1, \ldots, p\}$, we set $X_{\bf u}=(X_i, i \in \bf u)$ and $X_{\sim \bf u}=(X_i, i \in \{1,\ldots,p\}\setminus \bf u)$. 

\subsection{Definition of the generalized Sobol indices}
We recall the Hoeffding decomposition of $f$ \cite{van2000asymptotic}:
\begin{equation}
\label{e:hoeff}
f(X_1, \ldots, X_p) = c + f_{\bf u}(\Xu) + f_{\sim\bf u}(\Xtu) + f_{\bf u,\sim\bf u}(\Xu,\Xtu), 
\end{equation}
where $c\in\R^k$, $f_{\bf u}: \R^r\rightarrow\R^k$, $f_{\sim\bf u}: \R^{p-r}\rightarrow\R^k$ and $f_{\bf u,\sim\bf u}: \R^p\rightarrow\R^k$ are given by:
\[ c = \E(Y), \; f_{\bf u}=\E(Y|\Xu)-c, \; f_{\sim\bf u}=\E(Y|\Xtu)-c, \; f_{u,\sim\bf u}=Y-f_{\bf u}-f_{\sim\bf u}-c. \]
Taking the covariance matrices of both sides of \eqref{e:hoeff} gives (thanks to $L^2$-orthogonality):
\begin{equation}
\label{e:hoeffvar}
\Sigma = C_{\bf u} + C_{\sim\bf u} + C_{\bf u,\sim\bf u},
\end{equation}
where $\Sigma$, $C_{\bf u}$, $C_{\sim\bf u}$ and $C_{\bf u,\sim\bf u}$ are, respectively, the covariance matrices of $Y$, $f_{\bf u}(\Xu)$, $f_{\sim\bf u}(\Xtu)$ and $f_{\bf u,\sim\bf u}(\Xu,\Xtu)$.

For scalar outputs (ie., when $k=1$), the covariance matrices are scalar (variances), and \eqref{e:hoeffvar} is interpreted as the decomposition of the total variance of $Y$ as a sum of the variance caused by the variation of the input factors $X_i$ for $i\in\bf u$, the variance caused by the input factors not in $\bf u$, and the variance caused by the interactions of the factors in $\bf u$ and those not in $\bf u$. The (univariate) closed Sobol index $ S^{\bf u, \textrm{Scal}}(f) = \frac{C_{\bf u}}{\Sigma} $
is then interpreted as the sensibility of $Y$ to the inputs in $\bf u$. Due to noncommutativity of the matrix product, a direct generalization of this index is not straightforward.

We now go back to the general case. For any $k \times k$ matrix $M$, \eqref{e:hoeffvar} can be projected on a scalar by multiplying by $M$ and taking the trace:
\[ \Tr(M\Sigma) = \Tr(M \Cu) + \Tr(M \Ctu) + \Tr(M \Cutu). \] This equation is the natural scalarization of the matricial identity \eqref{e:hoeffvar} (as, for a symmetric matrix $V$, we have $\sum_{i,j} M_{i,j} V_{i,j} = \Tr(MV)$).  This suggests to define, when $\Tr(M \Sigma)\neq 0$: \[ S^{\bf u}(M; f) = \frac{\Tr(M \Cu)}{\Tr(M \Sigma)} \]
as the $M$-sensitivity measure (sensitivity index, or generalized Sobol index) of $Y$ to the inputs in $\bf u$. We can also analogously define:
$ S^{\sim\bf u}(M; f) = \frac{\Tr(M \Ctu)}{\Tr(M \Sigma)},\;\; S^{\bf u,\sim\bf u}(M; f) = \frac{\Tr(M \Cutu)}{\Tr(M \Sigma)}$, 
which measures the sensitivity to, respectively, the inputs not in $\bf u$, and to the interaction between inputs of $\bf u$ and inputs of $\{1,\ldots,p\}\setminus\bf u$.  The following lemma is obvious:
\begin{lemma}
\begin{enumerate}
\item The generalized sensitivity measures sum up to 1:
\begin{equation}
\label{e:somsobol}
S^{\bf u}(M; f) + S^{\sim \bf u}(M; f) + S^{\bf u, \sim \bf u}(M; f) = 1. 
\end{equation}
\item Left-composing $f$ by a linear operator $O$ of $\R^k$ changes the sensitivity measure according to:
\begin{equation}
\label{e:changesobol}
S^{\bf u}(M; Of) = \frac{\Tr(M O \Cu O^t)}{\Tr(M O \Sigma O^t)} = 
\frac{\Tr(O^t M O \Cu)}{\Tr(O^t M O \Sigma)} = 
S^{\bf u}(O^t M O; f).
\end{equation}
\item For $k=1$, and for any $M \neq 0$, we have $S^{\bf u}(M; f) = S^{\bf u,\textrm{Scal}}(f)$.
\end{enumerate}
\end{lemma}

\subsection{The case $M=\textrm{Id}_k$}
We now consider the special case $M = \textrm{Id}_k$ (the identity matrix of dimension $k$). We set $S_{\bf u}(f)=S^{\bf u}(\textrm{Id}_k;f)$. The index $S_{\bf u}(f)$ has the following properties:

\begin{e-proposition}
\label{prop:properties}
Suppose that $Y\in L^2(\Omega,\R^k)$ and that $\Sigma$ is positive-definite:
\begin{enumerate}
\item $0 \leq S_{\bf u}(f) \leq 1$;
\item $S_{\bf u}(f)$ is invariant by left-composition of $f$ by any isometry of $\R^k$, i.e.
\[ \forall O\, k\times k \text{ matrix s.t. } O^tO=\textrm{Id}_k, \;\; S_{\bf u}(Of)=S_{\bf u}(f); \]
\item $S_{\bf u}(f)$ is invariant by left-composition of $f$ by any nonzero homothety of $\R^k$.
\end{enumerate}
\end{e-proposition}

\emph{Proof: } Point (i): positivity is clear, as $\Cu$ and $\Sigma$ are positive; $S_{\bf u}(f)\leq1$ follows from positivity and \eqref{e:somsobol}. For (ii), we use \eqref{e:changesobol}. Point (iii) is immediate. \qed

The properties in the Proposition above are natural requirements for a sensitivity measure (the isometry invariance property ensures that the resulting indices are ``intrinsic'' and does not depend on the parametrization of the output). Note that $S_{\bf u}(f)$ is the sum of the partial variances divided by the sum of the total variances of each output coordinate, and the covariances between coordinates are not involved. In the next section, we will show that these requirements can be fulfilled by $S^{\bf u}(M;\cdot)$ iff $M=\lambda \textrm{Id}_k$ for $\lambda\in\R^*$. Hence the only ``canonical'' sensitivity measure is $S_{\bf u}$.

\section{$M=\textrm{Id}_k$ is the only good choice}
The following Proposition can be seen as a kind of converse of Proposition \ref{prop:properties}.

\begin{e-proposition}
Let $M$ be a square $k\times k$ matrix such that $\Tr(M V) \neq 0$ for any symmetric positive-definite matrix $V$. Now if for all $f: \R^p \rightarrow \R^k$, and all subsets $u\subset\{1,\ldots,p\}$, we have that 
$S^{\bf u}(M;f)$ is invariant by left-composition of $f$ by any isometry of $\R^k$, then $S^{\bf u}(M;\cdot) = S_{\bf u}(\cdot)$.
\end{e-proposition}

\emph{Proof:} Let $M$ as in the Proposition. We can write $M = M_{Sym} + M_{Antisym}$ where $M_{Sym}^t=M_{Sym}$ and $M_{Antisym}^t=-M_{Antisym}$. Since, for any symmetric matrix $V$, we have $\Tr(M_{Antisym} V)=0$, we have $S^{\bf u}(M;f)=S^{\bf u}(M_{Sym};f)$ and we can assume, without loss of generality, that $M$ is symmetric.

We diagonalize $M$ in an orthonormed basis: $M=P D P^t$, where $P^t P = \textrm{Id}_k$ and $D$ diagonal. We have:
\[ S^{\bf u}(M;f) = \frac{\Tr(P D P^t \Cu)}{\Tr(P D P^t \Sigma)} = \frac{\Tr(D P^t \Cu P)}{\Tr(D P^t \Sigma P)} = S^{\bf u}(D; P^tf). \]
This %(thanks to hypothesis (i)) 
shows that $M$ can in fact be assumed diagonal.

Now we want to show that $M = \lambda \textrm{Id}_k$ for some $\lambda\in\R^*$. Suppose, by contradiction, that $M$ has two different diagonal coefficients $\lambda_1 \neq \lambda_2$. It is clearly sufficient to consider the case $k=2$. Choose $f=\textrm{Id}_2$ (hence, $p=2$), and ${\bf u}=\{1\}$% (these choices are valid by (i) and (ii))
. We have $\Sigma=\textrm{Id}_2$, and $\Cu=\left(\begin{smallmatrix} 1 & 0 \\ 0 & 0 \end{smallmatrix}\right)$, hence on the one hand $S^{\bf u}(M;f)=\frac{\lambda_1}{\lambda_1+\lambda_2}$. On the other hand, let $O$ be the isometry which exchanges the two vectors of the canonical basis of $\R^2$. We have  $S^{\bf u}(M;Of)=\frac{\lambda_2}{\lambda_1+\lambda_2}$, and invariance by isometry is contradicted if % hence (iii) is contradicted 
$\lambda_1\neq\lambda_2$. We also have $\lambda \neq 0$ since $\Tr(M)\neq 0$. Finally, it is easy to check that, for any $\lambda\in\R^*$, $S^{\bf u}(\lambda \textrm{Id}_k;\cdot)=S^{\bf u}(\textrm{Id}_k;\cdot)=S_{\bf u}$. \qed

\section{Estimation of $S^{\bf u}(f)$}

In general, the covariance matrices $\Cu$ and $\Sigma$ are not analytically available. In the scalar case ($k=1$), it is customary to estimate $ S^{\bf u, \textrm{Scal}}(f) $ by using a Monte-Carlo pick-freeze method \cite{sobol1993,janon2012asymptotic}, which uses a finite sample of evaluations of $f$. In this Section, we propose a pick-freeze estimator for the vector case which generalizes the $T_N$ estimator studied in \cite{janon2012asymptotic}. We set: $Y^{\bf u}=f(X_{\bf u}, X_{\sim\bf u}')$ where $X_{\sim\bf u}'$ is an independent copy of $X_{\sim\bf u}$. Let $N$ be an integer. We take $N$ independent copies $Y_1, \ldots, Y_N$ (resp. $Y_1^{\bf u},\ldots,Y_N^{\bf u}$) of $Y$ (resp. $Y^{\bf u}$). For $l=1,\ldots,k$, and $i=1,\ldots,N$, we also denote by $Y_{i, l}$ (resp. $Y_{i, l}^{\bf u}$) the $l^\text{th}$ component of $Y_i$ (resp. $Y_i^{\bf u}$). We define the following estimator of $S_{\bf u}(f)$:
\[ S_{{\bf u}, N}(f) = \frac{\sum_{l=1}^k \left( \sum_{i=1}^N Y_{i,l} Y_{i,l}^{\bf u} - \frac1N \left( \sum_{i=1}^N \frac{Y_{i,l}+Y_{i,l}^{\bf u}}{2} \right)^2 \right)    }{ \sum_{l=1}^k \left( \sum_{i=1}^N \frac{ Y_{i,l}^2+(Y_{i,l}^{\bf u})^2 }{2} - \frac1N \left( \sum_{i=1}^N \frac{ Y_{i,l}+Y_{i,l}^{\bf u} }{2} \right)^2  \right) }.   \]
Thanks to the simple form of this estimator, the following Proposition can be proved in a way similar to the one used to prove Proposition 2.2 and Proposition 2.5 of \cite{janon2012asymptotic} (ie., by an application of the so-called Delta method).

\begin{e-proposition}
Suppose $Y \in L^4(\Omega,\R^k)$, and that $\Sigma$ is positive-definite. Then:
\begin{enumerate}
\item $\left(S_{{\bf u},N}(f)\right)_{N}$ is asymptotically normal: there exists $\sigma=\sigma(f)$ so that
$ \sqrt N ( S_{{\bf u},N}(f) - S_{\bf u}(f) ) $ converges (for $N \rightarrow +\infty$) in distribution to a centered Gaussian distribution with variance $\sigma^2$.
\item $\left(S_{{\bf u},N}(f)\right)_{N}$ is asymptotically efficient for estimating $S_{\bf u}(f)$ among regular estimator sequences that are function of exchangeable pairs $(Y, Y^{\bf u})$.
\end{enumerate}
\end{e-proposition}

{\scriptsize \textbf{Acknowledgements.~}
This work has been partially supported by the French National
Research Agency (ANR) through COSINUS program (project COSTA-BRAVA
n°ANR-09-COSI-015). }

\bibliographystyle{plain}
\bibliography{biblio}

\end{document}